\documentclass[10pt,conference]{IEEEtran}
\usepackage{amsmath,amssymb}
\usepackage{graphicx}
\usepackage{booktabs}
\usepackage{multirow}
\usepackage{cite}
\usepackage{url}
\usepackage{xcolor}
\usepackage{threeparttable}

\title{Aligning Quantum Operators with \\ Large Language Models}

\author{
\IEEEauthorblockN{Rogerio Feris\IEEEauthorrefmark{1}, Yunchao Liu\IEEEauthorrefmark{2}, Pengyuan Li\IEEEauthorrefmark{1}, Hang Hua\IEEEauthorrefmark{1}, and David Kremer\IEEEauthorrefmark{2}}
\IEEEauthorblockA{\IEEEauthorrefmark{1}MIT-IBM Computing Research Lab \IEEEauthorrefmark{2} IBM Quantum
\\
IBM Research 
}
}

\begin{document}
\maketitle

\begin{abstract}
Can Large Language Models (LLMs) understand and reason about quantum operators? Despite their remarkable capabilities in mathematics and symbolic reasoning, LLMs remain inherently blind to quantum representations such as unitary matrices. In this work, we take a step toward bridging this gap by introducing an approach that maps unitary operators into the latent space of an LLM, enabling unified modeling over quantum and linguistic inputs. We instantiate this idea on Clifford+T circuit synthesis over a Pauli rotation gate set, where our model achieves results competitive with state-of-the-art methods and scales consistently with training data, with no signs of saturation. Our approach further enables language-conditioned synthesis, allowing gate constraints unseen during training to be specified directly in natural language. This work suggests a path toward quantum--aware foundation models that can natively interpret and reason about quantum operations, which could have broader implications reaching across quantum compilation and algorithm discovery.
\end{abstract}

\section{Introduction}
\label{sec:intro}

Large language models have made great strides in recent years, demonstrating broad capabilities from code generation and competition-level mathematics to multimodal reasoning over images, audio, and structured data. These successes have catalyzed a few emerging efforts to bring LLMs into quantum computing, including Qiskit code assistant models \cite{dupuis2025quantum,ibm_granite_qiskit}, educational tools \cite{d2024exploring}, and multi-agent LLMs for OpenQASM programming \cite{fu2025qagent}.

Yet all of these systems share a common blind spot: they operate exclusively on symbolic representations of quantum objects, such as gate names, circuit descriptions, or quantum programs written as text. No existing approach equips an LLM to process the mathematical objects that actually define quantum operations, such as unitary matrices with complex-valued numerical structure. Current LLMs have no mechanism to ingest, interpret, or reason over such representations, while many tasks central to quantum compilation, verification, and algorithm design require direct access to the operator itself, not merely a human-readable label for it.

In this work, we take a step toward bridging this gap.  We introduce an alignment approach that projects unitary operators, represented as real-valued Pauli Transfer Matrices (PTM), into the latent space of a pretrained LLM via a lightweight encoder and projector, inspired by recent vision-language model architectures~\cite{team2025granite,liu2023visual}.  The resulting model accepts a quantum operator as a ``visual'' input alongside textual context, translates it into word embeddings understandable by the LLM, and autoregressively generates the output. To the best of our knowledge, this is the first approach that enables an LLM to condition directly on quantum operators rather than their textual or programmatic descriptions.


We validate our idea on the problem of unitary synthesis for 4-qubit operators, where the goal is to map a unitary operator to a circuit that implements it. We study unitary synthesis in a Pauli-rotation gate set, where each non-identity gate $R(P)=e^{-i\pi/8 P}$ is the Clifford-conjugate of a $T$ gate. This makes the setting closely connected to Clifford+$T$ circuits, while giving a uniform 256-way action space over Pauli strings.  We note that the PTM representation scales as $4^n \times 4^n$, which, like all full-unitary synthesis approaches, limits direct application to small qubit counts. However, the multimodal alignment framework itself is representation-agnostic: other quantum objects (e.g. Clifford tableaux, Pauli operator lists, or tensor-network descriptions) can be projected into the same LLM embedding space through additional modality-specific encoders, enabling a modular path toward larger-scale quantum compilation.

Training data is generated synthetically, exploiting the
asymmetry that computing a unitary on a few qubits is easy while
the inverse problem is already hard. At each synthesis step, the residual PTM (the portion of the target that remains to be compiled) is partitioned into non-overlapping patches (visual tokens), encoded by a lightweight encoder, and projected into the LLM's word embedding space via an MLP.  These visual tokens are concatenated with text embeddings carrying contextual information and an instruction prompt, and the LLM is fine-tuned to predict the next gate.

Our experiments show strong results that scale well with data, with no signs of a performance plateau, suggesting that further data would yield continued improvement.  On 4-qubit Clifford+$T$ synthesis with 1-15 gates, success rate improves more than $3\times$ as training data grows from 145K to 9.2M circuits.  Scaling at inference time (Best-of-N sampling) further boosts performance, surpassing a simulated-annealing baseline~\cite{paradis2024synthetiq} and prior reinforcement learning approaches~\cite{rietsch2024unitary}.
Beyond raw synthesis performance, we show that grounding synthesis in a language model unlocks a capability unavailable to specialized solvers: \emph{language-conditioned} synthesis, where the same model can be steered at inference time by natural language instructions. We demonstrate this capability  on a simple experiment with gate-set constraints unseen during training, showing the flexibility of our model.



More broadly, our results suggest a path toward quantum--aware multimodal models: systems that unify natural language and quantum representations within a shared embedding space. The alignment framework demonstrated here for PTMs can in principle accommodate additional quantum modalities, such as Clifford tableaux, Pauli operator lists, tensor-network descriptions, through modality-specific encoders, each projecting into the same LLM token space. Such models could leverage modern LLM capabilities (in-context learning, instruction following, multi-task transfer) for quantum compilation, transpilation, and beyond.  We view the present work as a proof of concept for this direction and plan to release our model and code publicly to support further research.

\section{Related Work}
\label{sec:related}


{\em LLMs and Quantum Computing.} Work at the intersection of quantum computing and large language models remains sparse, with only a handful of recent efforts bridging the two fields. On the code generation side, efforts include Granite for Qiskit~\cite{ibm_granite_qiskit}, Qiskit HumanEval~\cite{vishwakarma2024qiskit}, and KetGPT~\cite{apak2024ketgpt} for OpenQASM circuit generation. 
Beyond code generation, Agent-Q~\cite{jern2025agent} fine-tunes an LLM for circuit synthesis from text and graph inputs, QUASAR~\cite{yu2025quasar} extends this with an RL-based agentic pipeline, and \emph{QuantumLLMInstruct}~\cite{kashani2024quantumllminstruct}  provides a 500K instruction-tuning dataset for quantum reasoning. 
We depart from these approaches by enabling direct conditioning on  quantum representations: rather than operating on symbolic proxies, we map unitary operators directly into the LLM's latent space, allowing tasks such as text-conditioned unitary compilation and beyond.


\vspace{0.1in}
{\em Machine Learning-based Quantum Circuit Synthesis.} Classical approaches to exact Clifford+$T$ synthesis~\cite{kliuchnikov2013synthesis,matsumoto2008representation} and approximate single-qubit synthesis~\cite{ross2016optimal,kliuchnikov2016practical} provide optimality guarantees but scale poorly to multi-qubit unitaries.  The Solovay--Kitaev theorem~\cite{dawson2005solovay} guarantees that any single-qubit unitary can be approximated to precision $\epsilon$ with $O(\log^c(1/\epsilon))$ gates, and modern algorithms such as gridsynth~\cite{ross2016optimal} achieve near-optimal gate counts in practice.  However, extending these methods to multi-qubit synthesis remains an open challenge, motivating heuristic and learning-based alternatives.

Recent reinforcement learning (RL) methods have shown promise: Rietsch \emph{et al.}~\cite{rietsch2024unitary} synthesize Clifford+T circuits from unitary specifications, while Kremer \emph{et al.}~\cite{kremer2024practical} demonstrate practical RL-based synthesis for hardware-constrained settings and introduce in \cite{kremer2025optimizing} the Pauli-rotation basis and PTM representation for RL-based synthesis, using it to optimize non-Clifford gate counts.
AlphaTensor-Quantum~\cite{ruiz2025quantum} applies deep RL to minimize $T$-count in quantum circuits.  Beyond RL, GenQC~\cite{furrutter2024quantum} uses diffusion models to generate quantum circuits conditioned on desired properties, offering an alternative generative paradigm.

While promising, RL-based approaches require careful reward shaping, extensive hyperparameter tuning, and large amounts of environment interaction.  Our approach is \emph{RL-free}: we use only supervised fine-tuning with a standard next-token prediction loss.   

\section{Preliminaries: Quantum Circuit Synthesis}
\label{sec:preliminaries}

\begin{figure*}[t]
    \centering
    \includegraphics[width=0.8\textwidth]{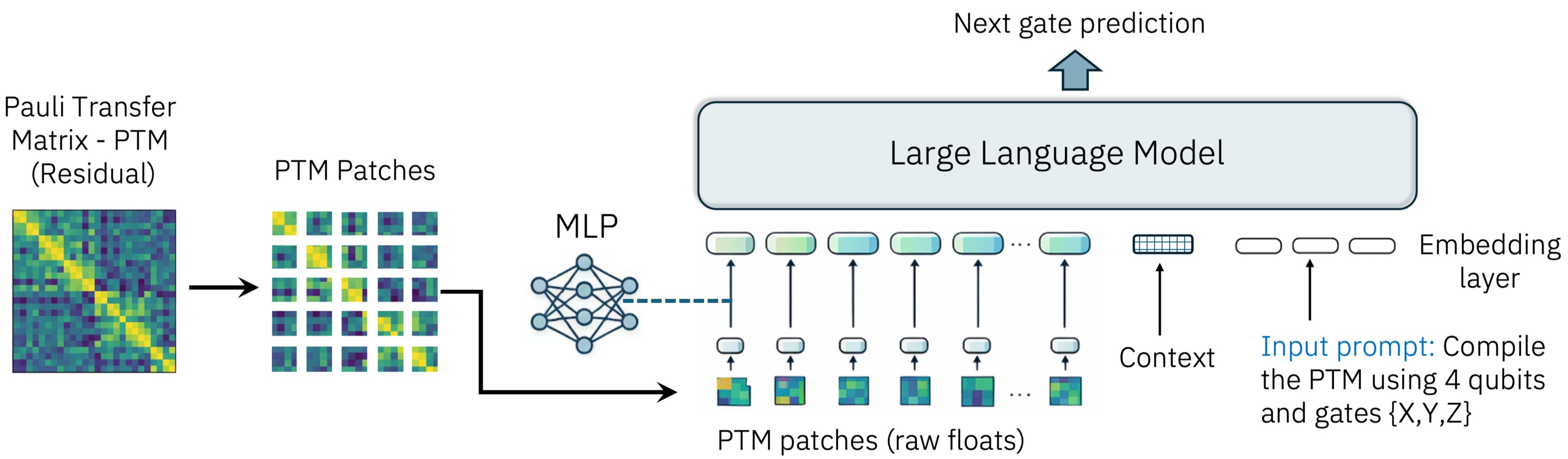}
    \caption{Overview of our approach. A target unitary $U$ is converted to its Pauli      Transfer Matrix (PTM), which is partitioned into non-overlapping patches and
  encoded by a lightweight encoder. An MLP projector maps the resulting  
  patch embeddings into the token space of a pretrained LLM, where they are   
  concatenated with text embeddings from the context (current fidelity and previous gates) and an instruction prompt. The LLM         
  autoregressively generates a sequence of Pauli rotation gates (e.g., IYYZ,    
  ZIII) that compile the target unitary. At each synthesis step, the residual
  PTM is re-encoded and fed back as a new visual input, enabling the model to
  condition each gate prediction on the remaining work.}
  \vspace{-0.1in}
    \label{fig:approach}
\end{figure*}

\subsection{Quantum circuit synthesis.}
An $n$-qubit quantum operation is described by a $d \times d$ unitary matrix $U \in \mathrm{U}(d)$, where $d = 2^n$.  \emph{Quantum circuit synthesis} is the problem of decomposing a target unitary~$U$ into a sequence of elementary gates from a fixed gate set~$\mathcal{G}$: find $(g_1, g_2, \dots, g_K)$ with $g_t \in \mathcal{G}$ such that $g_K \cdots g_2 \, g_1 \approx U$, where $K$ should be as small as possible.  The search space grows as $|\mathcal{G}|^K$, making exhaustive search intractable for all but very short circuits.

\subsection{Clifford+$T$ synthesis in a Pauli-rotation basis.}
The Clifford+$T$ gate set is the standard choice for fault-tolerant quantum computation.  Clifford gates alone can be efficiently simulated classically, but adding the $T$ gate ($\pi/8$ phase gate) yields a \emph{universal} set where any unitary can be approximated to arbitrary precision.  

Following~\cite{kremer2025optimizing}, we parameterize the gate set using $n$-qubit Pauli rotation gates.  The $n$-qubit Pauli group consists of $4^n$ operators $\{P_k\}$ formed by tensor products over $\{I, X, Y, Z\}$.  Each gate is a $\pi/8$-rotation:
\begin{equation}
\label{eq:pauli_rotation}
    R(P_k) = e^{-i\pi/8 \cdot P_k} = \cos\!\tfrac{\pi}{8}\, I - i\sin\!\tfrac{\pi}{8}\, P_k.
\end{equation}
For $n = 4$ qubits this gives $|\mathcal{G}| = 256$ gates, each labeled by a four-character string (e.g., \texttt{XIIX}). The circuit lengths reported in this paper are measured in this Pauli-rotation gate set. These rotations give a universal gate set; since they do not commute in general, gate order matters. This is closely related to the standard Clifford+$T$ gate set because each $R(P)$ (for a nonidentity $P$) can be viewed as rotating a single-qubit $T$ gate by an $n$-qubit Clifford unitary.

\subsection{Pauli Transfer Matrix (PTM).}
Rather than working with complex unitary matrices, we use the Pauli Transfer Matrix representation.  The PTM of a unitary~$U$ is a real-valued $4^n \times 4^n$ matrix:
\begin{equation}
\label{eq:ptm_def}
    \mathcal{P}_{ij} = \frac{1}{d}\,\mathrm{Tr}\!\bigl(P_i\, U\, P_j\, U^\dagger\bigr).
\end{equation}
The PTM is real-valued, invariant to global phase, and composes multiplicatively: the PTM of a circuit is the product of its gates' PTMs.  This last property is central to our stepwise approach.

\subsection{Fidelity.}
We measure synthesis progress via the channel fidelity between the residual PTM and the identity:
\begin{equation}
\label{eq:fidelity}
    \mathcal{F}(\mathcal{P}) = \mathrm{Tr}(\mathcal{P}) / 4^n.
\end{equation}
This equals~1 if and only if the target has been exactly synthesized.  We deem synthesis successful when $\mathcal{F} \geq 0.999$.

\section{Approach}
\label{sec:approach}

We present a multimodal alignment framework that maps quantum unitary operators into the latent space of a pretrained LLM, enabling autoregressive circuit compilation.  Given a target unitary, the model iteratively predicts Pauli rotation gates whose composition approximates it to high fidelity.  Our approach has two components: (i) a ~lightweight encoder and projector that aligns the PTM encoding with the LLM embedding space, and (ii)~a stepwise synthesis procedure. Figure \ref{fig:approach} shows an overview of our model architecture.

\subsection{PTM-Language Alignment}
\label{sec:alignment}

For $n{=}4$ qubits, the PTM (Eq.~\ref{eq:ptm_def}) is a $256 \times 256$ real matrix.  Before encoding, we normalize it element-wise to $[-1, 1]$ by dividing by its maximum absolute value, yielding a bounded input suitable for neural processing.
We treat the normalized PTM as a single-channel ``image'' and embed it into the LLM's token space via a lightweight encoder followed by an MLP projector.

\paragraph{PTM encoder}  The $256 \times 256$ PTM is partitioned into $16 \times 16$ non-overlapping patches, yielding $V = 256$ patch vectors of dimension~256.  Each patch is projected to a hidden dimension~$h_v = 768$ via a linear layer, followed by layer normalization and a learned positional embedding:
\begin{equation}
    \mathbf{z}_j = \mathrm{LayerNorm}\!\bigl(\mathbf{W}_\text{patch}\,\mathbf{p}_j\bigr) + \mathbf{e}_j, \quad j = 1, \dots, V.
\end{equation}

\paragraph{MLP projector}  A two-layer MLP with GELU activation maps the PTM embeddings to the LLM dimension~$d_\text{LLM}$:
\begin{equation}
    \mathbf{v}_j = \mathbf{W}_2\,\mathrm{GELU}\!\bigl(\mathbf{W}_1\,\mathbf{z}_j\bigr),
\end{equation}
where $\mathbf{W}_1 \in \mathbb{R}^{4h_v \times h_v}$ and $\mathbf{W}_2 \in \mathbb{R}^{d_\text{LLM} \times 4h_v}$.  The $V{=}256$ visual tokens are prepended to the text token embeddings, forming the input $[\mathbf{v}_1, \dots, \mathbf{v}_V, \mathbf{t}_1, \dots, \mathbf{t}_L]$ processed by the LLM.  The encoder and projector introduce ${\sim}14$M parameters ($<$0.4\% of total), leaving the LLM architecture unchanged.

\subsection{Stepwise Autoregressive Synthesis}
\label{sec:stepwise}

Rather than predicting the full gate sequence at
  once, we decompose compilation into a stepwise process: the model predicts one
   gate at a time, conditioned on the residual PTM, i.e. the portion of the target that remains to be synthesized.

Consider a target circuit $U = g_{T-1}\cdots g_1 g_0$, where $g_0$ acts first on the input state, and $g_k=R(P_k)$ for some Pauli operator $P_k$. Then the model predicts gates in reverse execution order, namely $g_{T-1}, g_{T-2}, \ldots, g_1, g_0$. The
  residual PTM is initialized as $\mathcal{P}^{(0)} = \operatorname{PTM}(U)$. At step $t\geq 0$, the model observes the residual PTM $\mathcal{P}^{(t)}$ and makes a prediction about the \emph{leftmost remaining factor} in the circuit decomposition of $\mathcal{P}^{(t)}$.
  
  After predicting a gate $\hat g$, the residual PTM $\mathcal{P}^{(t+1)}$ is updated by left-multiplying the inverse PTM of $\hat g$,
  $\mathcal{P}^{(t+1)} = \operatorname{PTM}(\hat g)^{-1} \mathcal{P}^{(t)} = \operatorname{PTM}(\hat g)^{\top} \mathcal{P}^{(t)}$. Ideally, as synthesis
  progresses, the residual PTM approaches the identity matrix and the fidelity between the residual PTM and the identity $\mathcal{F}(\mathcal{P}^{(t)}) = \mathrm{Tr}(\mathcal{P}^{(t)}) / 4^n$ increases toward 1.
  
  Iterating this peeling step yields the full prediction sequence $\hat g_{T_{\text{pred}}-1}, \hat g_{T_{\text{pred}}-2}, \ldots, \hat g_0$  (we index the predicted sequence so that $\hat g_{T_{\text{pred}}-1}$ is the first output of the model and $\hat g_0$ is the last), and the induced unitary is $\hat U = \hat g_{T_{\text{pred}}-1}\cdot\hat g_{T_{\text{pred}}-2}\cdots \hat g_0$. Note that the predicted sequence $\hat g_{T_{\text{pred}}-1}, \hat g_{T_{\text{pred}}-2}, \ldots, \hat g_0$ could be very different from the ground truth sequence $g_{T-1}\cdots g_1 g_0$ that is used to construct $U$, while the unitary $\hat U$ can be exactly equal or very close to $U$ (up to a global phase).

  

  More specifically, at each step the model receives the following input: 1) the residual PTM encoded as visual tokens; 2) the context information encoded as text, including the current fidelity and the previously predicted gates; 3) an instruction prefix that directs the large language model to compile the input unitary using a specified gate set.
  As output, the model generates a
  single gate prediction, after which the residual is updated externally and the
   process repeats. The vocabulary includes a special END token.
  This formulation reduces the combinatorial synthesis problem to a sequence of
  single-token decisions, each grounded in a visual representation of the
  remaining work. The external PTM update acts as a "scratchpad" — the model
  need not maintain synthesis state internally, as the residual is re-encoded at
   every step.

  \paragraph{Training objective.} Each training sample corresponds to a single
  synthesis step: given the current residual PTM $\mathcal{P}^{(t-1)}$, the     
  model must predict the next gate $a_t$. It is trained with the standard
   causal language modeling objective, predicting   
  the gate conditioned on the visual and textual inputs:                      
  \begin{equation}
      \mathcal{L} = -\log p_\theta(a_t \mid \mathbf{v}_1, \dots, \mathbf{v}_V,
  \mathbf{h}_1, \dots, \mathbf{h}_H, \mathbf{q}_1, \dots, \mathbf{q}_Q),
  \end{equation}
  where $\mathbf{v}_1, \dots, \mathbf{v}_V$ are vision token
  embeddings encoding $\mathcal{P}^{(t-1)}$, $\mathbf{h}_1, \dots, \mathbf{h}_H$
   are the context tokens encoding the current fidelity
  $\mathcal{F}(\mathcal{P}^{(t-1)})$ and the recently predicted gates, and
  $\mathbf{q}_1, \dots, \mathbf{q}_Q$ are the instruction tokens 
   directing the model to compile the input. The loss
  is computed exclusively over the gate prediction; vision, context, and
  instruction positions are masked and do not contribute gradients.

  Training samples are generated on the fly: at each iteration, a random
  Clifford+$T$ circuit of length $K$ is sampled (denoted as $U=g_{K-1}\cdots g_1 g_0$), and the $K{+}1$ stepwise
  decompositions (including the final \texttt{END} prediction when
  $\mathcal{F}(\mathcal{P}^{(K)}) \approx 1$) are used as training
  examples. The gates are presented in synthesis order, i.e. reverse execution order:
  $g_{K-1}, g_{K-2}, \ldots, g_0$.
  Equivalently, the model learns to peel off the leftmost remaining factor in the written product $U = g_{K-1}\cdots g_0$.


\begin{table}[t]
\centering
\begin{threeparttable}
\caption{Data scaling on 1--15 gate circuits.  All models use the same architecture and hyperparameters; only the number of training circuits varies.  Evaluated on 2{,}000 held-out circuits with fidelity threshold $\tau = 0.999$.}
\label{tab:data_scaling}
\begin{tabular}{lcc}
\toprule
\textbf{Training circuits} & \textbf{Success (\%)} & \textbf{Mean fidelity} \\
\midrule
145K  & 23.4 & 0.477 \\
287K  & 25.5 & 0.475 \\
575K  & 37.3 & 0.607 \\
1.15M & 58.1 & 0.658 \\
2.3M  & 62.9 & 0.688 \\
4.6M  & 66.7 & 0.711 \\
9.2M  & 71.0 & 0.746 \\
\midrule
9.2M $\rightarrow$ 4.6M (1--30 gates)\textsuperscript{$\dagger$} & 87.9 & 0.894 \\
\bottomrule
\end{tabular}
\begin{tablenotes}
\footnotesize
\item[$\dagger$] Initialized from the 9.2M checkpoint and continued training on additional 4.6M circuits spanning 1--30 gates. Evaluated on the same 1--15 gate held-out set.
\vspace{-0.1in}
\end{tablenotes}
\end{threeparttable}
\end{table}

\subsection{Two-Stage Training}
\label{sec:training}

\paragraph{Stage~1: Projector alignment.}  Only the vision encoder and projector are optimized; the LLM is frozen.  
This establishes cross-modal alignment without perturbing pretrained representations (${\sim}$7K steps, LR $10^{-3}$, cosine decay).

\paragraph{Stage~2: Joint fine-tuning.}  All parameters are jointly optimized with \emph{differential learning rates}: a lower rate~$\eta_\text{LLM}$ for the language model and a higher rate~$\eta_\text{proj} \approx 4\eta_\text{LLM}$ for the vision components.  We use a Warmup--Stable--Decay (WSD) schedule: linear warmup, constant for 75\% of training, then decay to 5\% of peak over the final 25\%.

\begin{table}[t]
\centering
\caption{Success rate (\%) by circuit length with best-of-$N$ sampling.  $N{=}1$: greedy decoding.  Evaluated on 2{,}000 held-out circuits with fidelity threshold $\tau = 0.999$.}
\label{tab:bon_results}
\setlength{\tabcolsep}{3pt}
\begin{tabular}{lrrrrrrrr}
\toprule
\textbf{Length} & $N{=}1$ & $N{=}3$ & $N{=}5$ & $N{=}7$ & $N{=}10$ & $N{=}20$ & $N{=}40$ & $N{=}80$ \\
\midrule
1--5  & 99.9 & 99.9 & 99.9 & 99.9 & 99.9 & 99.9 & 100.0 & 100.0 \\
6     & 98.5 &  99.2 & 100.0 & 100.0 & 100.0 & 100.0 & 100.0 & 100.0 \\
7     & 97.1 &  98.6 & 100.0 &  99.3 & 100.0 & 100.0 & 100.0 & 100.0 \\
8     & 95.6 &  97.8 &  99.3 &  99.3 & 100.0 & 100.0 & 100.0 & 100.0 \\
9     & 91.2 &  97.1 &  99.3 &  99.3 &  98.5 & 100.0 &  99.3 & 100.0 \\
10    & 83.3 &  92.1 &  96.5 &  98.2 &  99.1 & 100.0 & 100.0 & 100.0 \\
11    & 83.3 &  90.3 &  91.7 &  96.5 &  95.8 &  95.8 &  98.6 &  99.3 \\
12    & 75.0 &  85.8 &  88.5 &  93.2 &  93.9 &  98.0 &  98.0 &  99.3 \\
13    & 76.4 &  88.2 &  88.2 &  91.8 &  95.5 &  98.2 &  98.2 &  99.1 \\
14    & 59.5 &  69.0 &  77.0 &  84.9 &  87.3 &  87.3 &  92.1 &  97.6 \\
15    & 57.2 &  71.7 &  76.1 &  79.0 &  87.0 &  88.4 &  92.0 &  94.9 \\
\midrule
\textbf{All} & \textbf{87.9} & \textbf{92.7} & \textbf{94.5} & \textbf{96.1} & \textbf{97.1} & \textbf{97.8} & \textbf{98.6} & \textbf{99.4} \\
\bottomrule
\vspace{-0.3in}
\end{tabular}
\end{table}

\section{Experiments}
\label{sec:experiments}

We evaluate our approach on Clifford+$T$ synthesis of 4\nobreakdash-qubit unitaries.  Unless noted otherwise, we use Granite 4.0 Micro~\cite{granite40micro} (3B parameters) as the LLM backbone, train on 2 nodes $\times$ 8 GPUs with an effective batch size of~64, and evaluated on held-out circuits with verified zero overlap against all training splits, using a fidelity threshold of~$\tau = 0.999$.

\subsection{Circuit Synthesis Results: Data Scaling}
\label{sec:circuit_results}

We train models on circuits with 1--15 gates, varying only the number of training circuits from 145K to 9.2M while keeping all other hyperparameters fixed.  Table~\ref{tab:data_scaling} reports success rate and mean fidelity on 2{,}000 held-out circuits.
Performance improves consistently with more training data, achieving more than $3\times$ improvement in success rate from 145K to 9.2M circuits. We also explore scaling along the gate-count dimension: initializing from the 9.2M checkpoint and continuing training on additional 4.6M circuits spanning 1–30 gates. This model achieves 87.9\% greedy success on the same 1–15 gate held-out set, a gain of nearly 17 percentage points over the 9.2M baseline trained on 1–15 gates alone, indicating that exposure to longer circuits substantially improves synthesis for shorter ones.   Neither trend has yet saturated, suggesting that further scaling is possible.  

\vspace{-0.02in}
\subsection{Inference-Time Scaling}
\label{sec:inference_scaling}

Best-of-$N$ sampling provides a simple mechanism for trading inference-time compute for accuracy.  For each circuit, we run $N$ independent synthesis rollouts --- the first using greedy decoding and the remaining $N{-}1$ at temperature 0.7 --- and keep the result with the highest fidelity, with early termination once fidelity exceeds~$\tau$.
Table~\ref{tab:bon_results} shows per-length success rates for the 30-gate model with increasing~$N$.

With greedy decoding alone ($N{=}1$), the model achieves 87.9\% success.  Increasing to $N{=}10$ raises this to 97.1\%, and $N{=}80$ reaches 99.4\%, an 11.5-percentage-point improvement from sampling alone.  The gains are roughly log-linear in~$N$ and are concentrated on longer circuits (11--15 gates), where stochastic exploration can discover synthesis paths that greedy decoding misses.  This shows that the model learns a well-calibrated distribution over synthesis paths rather
  than collapsing to a single strategy: even when greedy decoding fails, the model assigns meaningful
  probability to correct alternatives that stochastic sampling can recover, making inference-time
  scaling an effective complement to training-time data scaling. 

\begin{figure}[t]
\centering
\includegraphics[width=1.0\columnwidth]{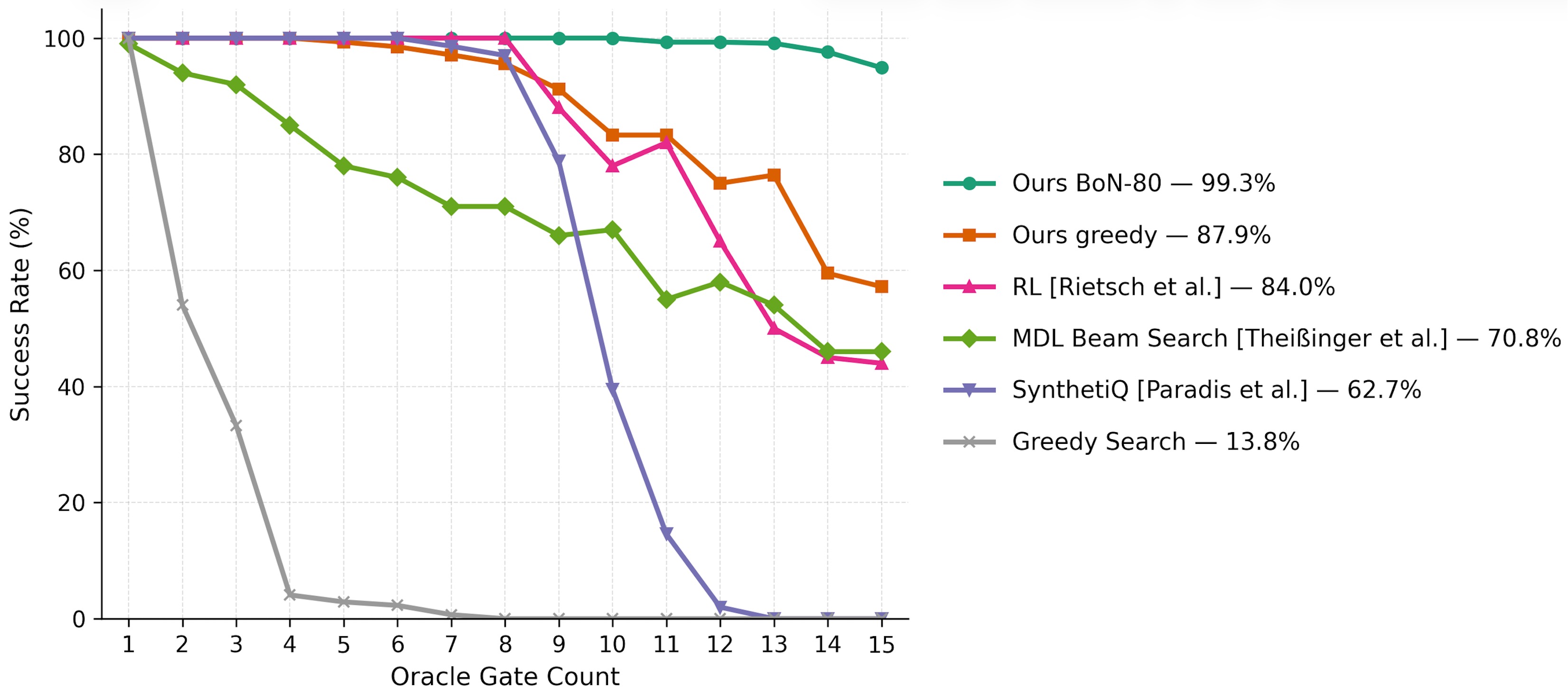}
\caption{Per-gate success rate on 2{,}000 held-out circuits (1--15 gates, fidelity threshold $\tau = 0.999$).  RL and MDL beam search results are taken  
  from their respective publications as approximate comparisons (see text). Overall weighted success rates shown in the legend.}
  \vspace{-0.15in}
\label{fig:baselines}
\end{figure}

\subsection{Baselines}
\label{sec:baselines}

Figure~\ref{fig:baselines} shows per-gate success rates on 2{,}000 held-out circuits (1--15 gates).  We compare against four baselines: greedy search (best-of-256 at each step), SynthetiQ~\cite{paradis2024synthetiq} (simulated annealing, 100\,s budget on 48 CPU threads), the RL approach of Rietsch et~al.~\cite{rietsch2024unitary} (Gumbel AlphaZero), and the MDL beam search of Thei{\ss}inger et~al.~\cite{theissinger2026beyond}.  For SynthetiQ we ran the publicly available code on our held-out set; we could not find code available for RL and MDL beam search methods, so we report their published results as approximate comparisons. Note that all methods target 4-qubit Clifford+$T$ synthesis, but the underlying unitary distributions may differ slightly: our circuits
are products of Pauli rotations while theirs are generated directly in Clifford+$T$.

Greedy search achieves only 13.8\% overall, collapsing beyond 3~gates.  SynthetiQ reaches 62.7\% with near-perfect synthesis up to 6~gates but falls to 0\% at 13~gates and above.  MDL beam search (68.8\%) starts below RL (83.7\%) on short circuits but surpasses it for longer ones, as its search strategy scales better with gate count.  RL achieves strong results on shorter circuits but degrades sharply beyond 13~gates.

Our model surpasses all baselines with 87.9\% success using greedy decoding alone.  With best-of-80 sampling, it reaches 99.4\% overall, maintaining above 94\% accuracy even at 15~gates where all other methods struggle.  Our model runs inference in approximately 1\,s per sample on a single NVIDIA H100 GPU, or roughly 80\,s for best-of-80. As a reference, MDL beam search reports 22\,s per sample.  For successful circuits, the model produces sequences with a mean predicted-to-oracle gate ratio of 1.007, indicating that it learns near-optimal-length decompositions rather than
trading accuracy for longer circuits. Notably, our approach does not require reward shaping, exploration strategies, or the extensive hyperparameter tuning typical of RL\@.  Moreover, RL methods such as GRPO can be applied on top of our SFT-trained model as a fine-tuning stage, potentially yielding further gains.

\subsection{Haar Random Unitaries}
\label{sec:haar}

Haar-random unitaries, sampled uniformly from the unitary group, are not in general expressible as finite Clifford+$T$ circuits and thus lie outside the training distribution.  Compiling such unitaries is the problem of \emph{approximate synthesis}: the Solovay--Kitaev theorem~\cite{dawson2005solovay} guarantees that any single-qubit unitary can be $\epsilon$-approximated with $O(\log^c(1/\epsilon))$ Clifford+$T$ gates, and modern algorithms~\cite{ross2016optimal} achieve near-optimal gate counts. Extending these guarantees to multi-qubit unitaries remains an open problem.  We do not expect exact synthesis, but use Haar-random unitaries as a stress test: can the model learn to make meaningful progress toward compiling arbitrary unitaries?

\begin{figure}[t]
    \centering
    \includegraphics[width=0.75\columnwidth]{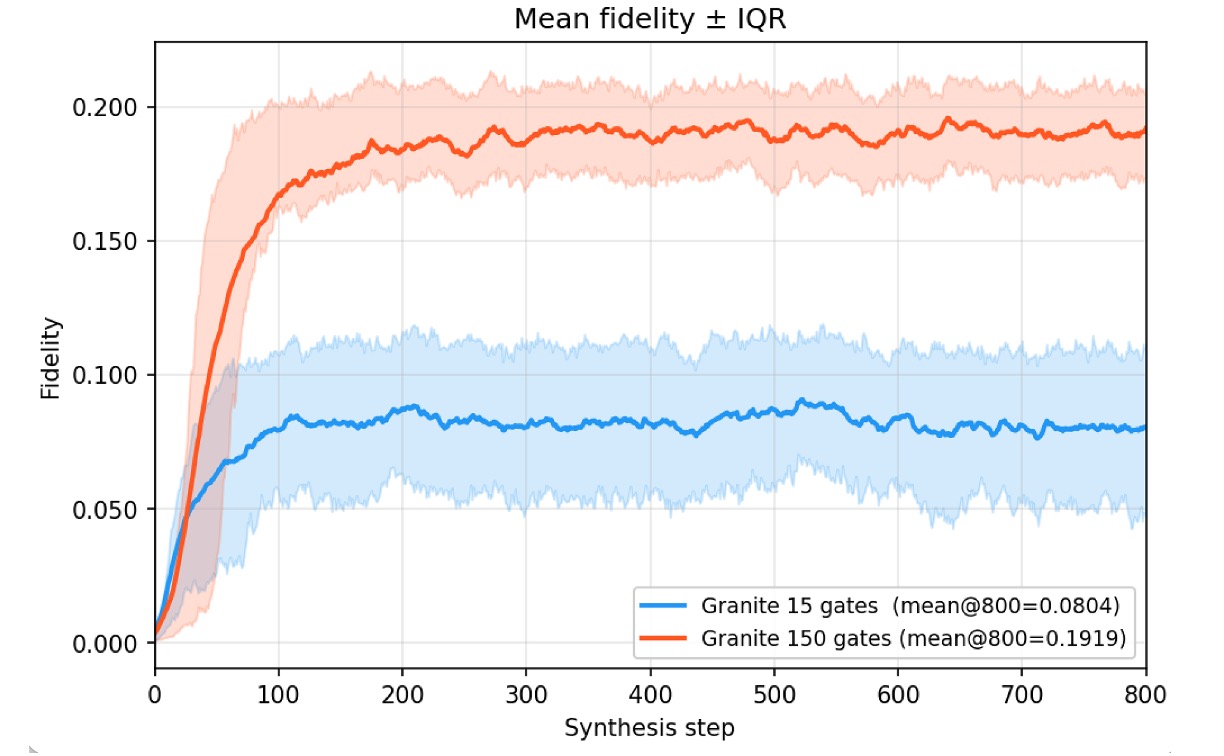}
    \vspace{-0.1in}
    \caption{Fidelity over 800 synthesis steps on 200 Haar-random 4-qubit unitaries. The 150g model achieves substantially higher fidelity than the 15g model, suggesting that training on longer circuits improves generalization to arbitrary unitaries.}
    \vspace{-0.2in}
    \label{fig:haar}
\end{figure}

Figure~\ref{fig:haar} shows fidelity progress over 800 synthesis steps on 200 Haar-random unitaries.  The model trained on 1--15 gate circuits makes limited progress, with mean fidelity plateauing below~0.02.  In contrast, we train a model on 1--150 gates with only 1M circuits, and show that it achieves substantially higher mean fidelity.  While these fidelities are far from the $\geq 0.999$ threshold needed for exact synthesis, the monotonic improvement with training gate range suggests that scaling to longer circuits is a potentially viable path toward compiling increasingly general unitaries.

\begin{figure*}[t]
    \centering
    \includegraphics[width=0.9\textwidth] {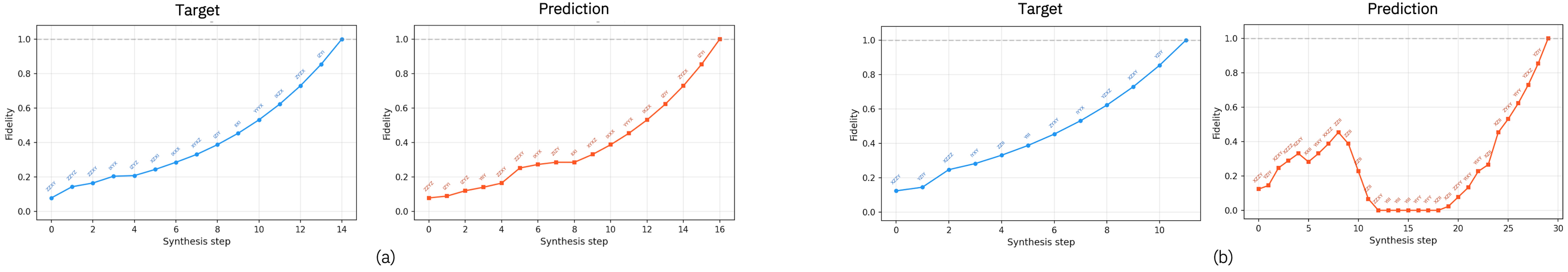}
    \vspace{-0.1in}
    \caption{Fidelity traces for training oracle sequences and model predictions at test time. The model's predictions generalize beyond the training data, most strikingly in~(b), where fidelity rises, then drops, and finally recovers to~1.0.}
    \label{fig:ushape}
    \vspace{-0.2in}
\end{figure*}

\begin{figure*}[t]
    \centering
    \includegraphics[width=0.9\textwidth]{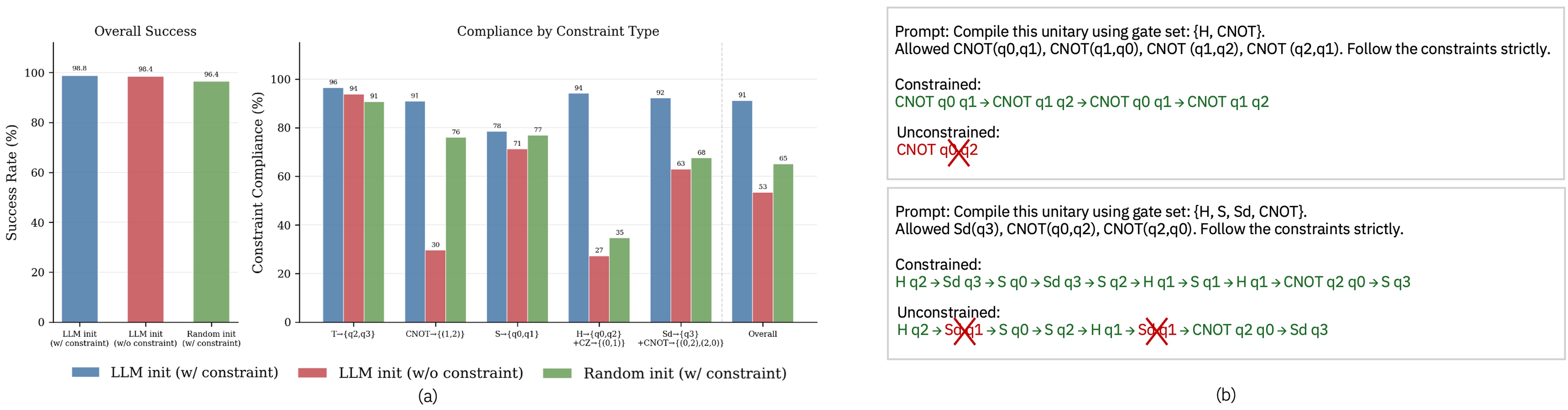}
    \vspace{-0.1in}
    \caption{Text-conditioned synthesis on unseen constraint configurations. (a)   
  Overall success rate (similar for the three settings) and gate-level constraint compliance per constraint type. The large drop when removing constraint text from
  the prompt confirms that the model actively conditions on
  the instruction rather than producing compliant outputs by default, while the
  gap between LLM and random initialization shows that
  pretrained language understanding is critical for interpreting unseen placement
  constraints. (b) Qualitative examples for constrained and unconstrained prompts.}
  \vspace{-0.2in}
    \label{fig:conditioned}
\end{figure*}

\subsection{Patch Size Ablation}
\label{sec:patch_ablation}

The patch size $P$ of our visual encoder controls the trade-off between spatial fidelity and sequence length: smaller patches preserve fine-grained
  PTM structure but increase the number of visual tokens the language model must
   attend to. We ablate this choice with $P \in {8, 16, 32, 64, 256}$,
  corresponding to ${1024, 256, 64, 16, 1}$ visual tokens. All models are       
  trained on 1.15M circuits (1--15 gates, 4 qubits) with identical              
  hyperparameters and evaluated on 2000 held-out circuits via greedy
  autoregressive synthesis. Patch sizes 8 and 16 achieve nearly identical overall success
  (60.1\% and 59.4\%), while performance degrades substantially at $P{=}32$       
  (39.5\%) and beyond (64: 39.5\%, 256: 31.4\%). We adopt $P{=}16$ for all other
  experiments.


\subsection{Qualitative Results}
\label{sec:qualitative}

Figures~\ref{fig:ushape}(a)--(b) show fidelity traces for example pairs of target circuit (oracle) and corresponding prediction at test time. They clearly show the model generalize beyond imitation, as these trajectories are entirely absent from the training distribution, suggesting the model has internalized the PTM structure rather than memorizing input--output mappings. Figure~\ref{fig:ushape}(b), in particular, shows an interesting case where fidelity initially rise, then drops sharply before the model reverses course and recovers to~1.0. Despite such cases, the mean predicted-to-oracle gate ratio is approximately one, indicating near-optimal-length decompositions overall.

\subsection{Text-Conditioned Circuit Synthesis}
\label{sec:conditioned}

A key advantage of grounding quantum circuit synthesis in a language model is the ability to condition generation on natural language instructions.  We explore this by training the model to follow \emph{gate-set constraints}: text prompts that restrict which qubits or qubit pairs a given gate may act on.  Such constraints arise naturally in hardware-aware compilation, where qubit connectivity limits the set of physically realizable two-qubit interactions, and in routing, where CNOT gates must be decomposed over adjacent pairs.

For this experiment, we use the gate set $\{$H, T, T$^\dagger$, S, S$^\dagger$, X, Y, Z, CNOT, CZ$\}$, with each training circuit composed of gates sampled randomly from this set. Each circuit is paired with a randomly sampled constraint configuration: with probability~0.5, the prompt specifies one or two restrictions on gate placement (e.g., ``Allowed T(q0, q2)''); otherwise, the prompt imposes no constraints. The model was trained with 3M circuits, 1-15 gates. It receives the constraint text as part of its input and must produce a gate sequence that both implements the target unitary and respects the specified restrictions. Compliance is measured at the gate level: for each restricted gate the model predicts, we check whether it respects the constraint.

To evaluate generalization, we construct an out-of-distribution test set of 250~circuits with five {\em unseen constraint combinations} (blacklisted during training), ranging from single-gate restrictions to dual constraints. We evaluate three settings: LLM init with constraint prompts, LLM init with unconstrained       
  prompts (identical gate sets but no restriction text), and random weights init with constraint prompts. All three achieve comparable synthesis success but constraint compliance differs sharply
  (Figure~\ref{fig:conditioned}). The LLM-initialized model achieves 91\% compliance, dropping to 53\% when the constraint text is removed, confirming that the model actively conditions on the instruction rather than producing compliant outputs by default. The randomly initialized model reaches only 65\%
  compliance despite receiving the same constraint prompts, demonstrating that pretrained language understanding is critical for interpreting unseen placement constraints.

These results show that our model is flexible and can synthesize circuits with or without constraints, steered by the input prompt. We believe the interplay between language and quantum operators opens directions beyond constraint following: explanations of synthesis choices, interactive debugging, and potentially unlocking chain-of-thought reasoning over quantum states, akin to the deliberative ``wait'' and ``aha'' moments observed in large language models~\cite{liu2024deepseek}.  We leave exploration of these directions to future work.

\vspace{-0.02in}
\section{Conclusion}
\label{sec:conclusion}

We have presented an approach that maps quantum unitary operators into the latent space of a large language model, enabling native reasoning over quantum representations.  Instantiated on Clifford+$T$ circuit synthesis for 4-qubit unitaries, our method achieves strong results -- including text-conditioned synthesis -- using only supervised fine-tuning, and performance scales consistently with both training data and inference-time compute without signs of saturation.

Our work opens several directions for future research. Natural next steps include scaling to larger qubit counts and deeper circuits, and augmenting training with (latent) reasoning and GRPO to further improve recovery and optimality, and incorporating different circuit and operator representations. Ultimately, we envision a class of quantum–language models that unify textual context, instruction following, and direct operator reasoning within a single system. We plan to release our model publicly to support this line of work.


\bibliographystyle{IEEEtran}
\bibliography{references}

\end{document}